\documentclass[article,aps,prl,reprint,twocolumn,groupedaddress,10pt]{revtex4-1}
\usepackage{amsmath}%,preprint
\usepackage{amsthm}
\usepackage{mathrsfs}
\usepackage{amsfonts}
\usepackage{amssymb}
\usepackage{graphicx}
\usepackage[colorlinks,linkcolor=blue,anchorcolor=blue,urlcolor=blue,citecolor=blue]{hyperref}
\usepackage{eufrak}
\usepackage{multirow}
\usepackage{textcomp}
\usepackage{epstopdf}
\usepackage{float}
\usepackage{epstopdf}

\begin{document}

\title{Experimental realization of universal high-dimensional quantum gates with ultra-high fidelity and efficiency}

\author{Zhe Meng}\thanks{These two authors contributed equally to this work.}
\affiliation{Center for Quantum Technology Research and Key Laboratory of Advanced Optoelectronic Quantum Architecture and Measurements (MOE), School of Physics, Beijing Institute of Technology, Beijing 100081, China}	
\author{Wen-Qiang Liu}\thanks{These two authors contributed equally to this work.}
\affiliation{Center for Quantum Technology Research and Key Laboratory of Advanced Optoelectronic Quantum Architecture and Measurements (MOE), School of Physics, Beijing Institute of Technology, Beijing 100081, China}	

\author{Bo-Wen Song}
\affiliation{Center for Quantum Technology Research and Key Laboratory of Advanced Optoelectronic Quantum Architecture and Measurements (MOE), School of Physics, Beijing Institute of Technology, Beijing 100081, China}	

\author{Xiao-Yun Wang}
\affiliation{Center for Quantum Technology Research and Key Laboratory of Advanced Optoelectronic Quantum Architecture and Measurements (MOE), School of Physics, Beijing Institute of Technology, Beijing 100081, China}

\author{An-Ning Zhang}
\email{Corresponding author. Anningzhang@bit.edu.cn}
\affiliation{Center for Quantum Technology Research and Key Laboratory of Advanced Optoelectronic Quantum Architecture and Measurements (MOE), School of Physics, Beijing Institute of Technology, Beijing 100081, China}	
\date{\today }

\author{Zhang-Qi Yin}
\email{Corresponding author. zqyin@bit.edu.cn}
\affiliation{Center for Quantum Technology Research and Key Laboratory of Advanced Optoelectronic Quantum Architecture and Measurements (MOE), School of Physics, Beijing Institute of Technology, Beijing 100081, China}

\begin{abstract}
Qudit, a high-dimensional quantum system, provides a larger Hilbert space to process the quantum information and has shown remarkable advantages over the qubit counterparts. 
It is a great challenge to realize the high fidelity universal quantum gates with qudits. 
Here we theoretically propose and experimentally demonstrate a set of universal quantum gates for a single optical qudit with four dimensions (including the generalized Pauli $X_4$ gate, Pauli $Z_4$ gate, and all of their integer powers), which are encoded in the polarization-spatial degree of freedom without multiple unstable cascaded interferometers. Furthermore, we also realize the controlled-$X_4$ gate and all of its integer powers. 
We have achieved  both the ultra-high average gate fidelity $99.73\%$ and efficiency $99.47\%$, which are above the the error threshold for fault-tolerant quantum computation.  Our work paves a way for the large-scale high-dimensional fault-tolerant quantum computation with a polynomial resource cost. 
\end{abstract}

%\pacs{03.67.Lx, 42.50.Ex, 05.70.Ln}

\maketitle

%\section{Introduction}

\emph{Introduction.}---Universal quantum logic gates are  essential building blocks in many quantum information processing tasks \cite{nielsen2010quantum}. Recently, the realization of qubit-based quantum gates in many physical platforms has been theoretically developed and experimentally demonstrated well \cite{knill2001scheme,o2003demonstration,zhao2005experimental,zeuner2018integrated,li2021heralded,liu2022universal,liu2023linear}.
In addition to qubit, qudit with $d$-ary ($d > 2$) digits has emerged as a richer resource in high-dimensional quantum systems and it has been extended to high-dimensional logic to encode and process quantum information \cite{luo2014universal,wang2020qudits}. Due to the larger operation space, qudit quantum systems have shown their remarkable advantages. For example,  simplifying quantum gates \cite{ralph2007efficient,lanyon2009simplifying,liu2020optimal,liu2020low}, improving the efficiency of fault-tolerant quantum computation \cite{campbell2014enhanced,howard2017application}, increasing channel capacity \cite{bechmann2000quantum,cortese2004holevo,dixon2012quantum,hu2018beating}, improving communication security \cite{cerf2002security,zhang2008secure,wang2020high}, etc. Besides, the qudits  can exceed the limitations imposed by the qubits in larger violation of Bell-type inequality \cite{collins2002bell,vertesi2010closing,dada2011experimental} and  higher noise resilience \cite{sheridan2010security,liu2009decay,ecker2019overcoming}. Up to now, qudit-based quantum information processing has been reported and experimentally implemented in various physical systems \cite{anderson2015accurate,erhard2018twisted,ringbauer2109universal,cozzolino2019high,erhard2020advances,blok2021quantum,cervera2022experimental,fu2022experimental,chi2023high}. Though high-dimensional quantum information processing has made significant progress, a lot of efforts still are required for improving the quantum gates fidelity \cite{erhard2018twisted,cozzolino2019high}.

Photon is a natural candidate for encoding the qudit due to its various degrees of freedom (DOFs).
There are many experiments that have been demonstrated universal quantum gates for the qudits, which are formed by
the orbital angular momentum (OAM) \cite{babazadeh2017high,gao2019arbitrary,gao2020computer,wang2021experimental}, the time-frequency DOF \cite{imany2019high}, or the spatial modes of photons \cite{brandt2020high,chi2022programmable}. In 2017,  Babazadeh \emph{et al.}  experimentally demonstrated a four-dimensional generalized Pauli $X_4$ gate and all of its integer powers with a conversion efficiency of 87.3\% and a fidelity of 93.4\% using the OAM mode of a single photon \cite{babazadeh2017high}. Later,  Wang \emph{et al.} improved the conversion efficiency to 93\%  \cite{wang2021experimental}. In 2022,  Chi \emph{et al.} experimentally realized the Pauli $X_4$ gate with the fidelity of 98.8\% and a four-dimensional controlled-$X_4$ gate with the fidelity of 95.2\% on a programmable silicon-photonic quantum processor \cite{chi2022programmable}. In all above works \cite{babazadeh2017high,gao2019arbitrary,gao2020computer,wang2021experimental,chi2022programmable}, the experimental realization of the high-dimensional quantum gates depends on either the multiple Sagnac--type interferometers or multiple Mach--Zehnder--type interferometers, which are both faced with the great challenges of phase instability. Therefore, the fidelity and efficiency of the universal quantum gates for the optical qudits are significantly degraded, compared to their qubit counterparts \cite{zeng2016realization,lopes2019linear,ke2024versatile}.

\begin{figure*}   %[htpb] %[!h]
\begin{center}
\includegraphics[width=17cm,angle=0]{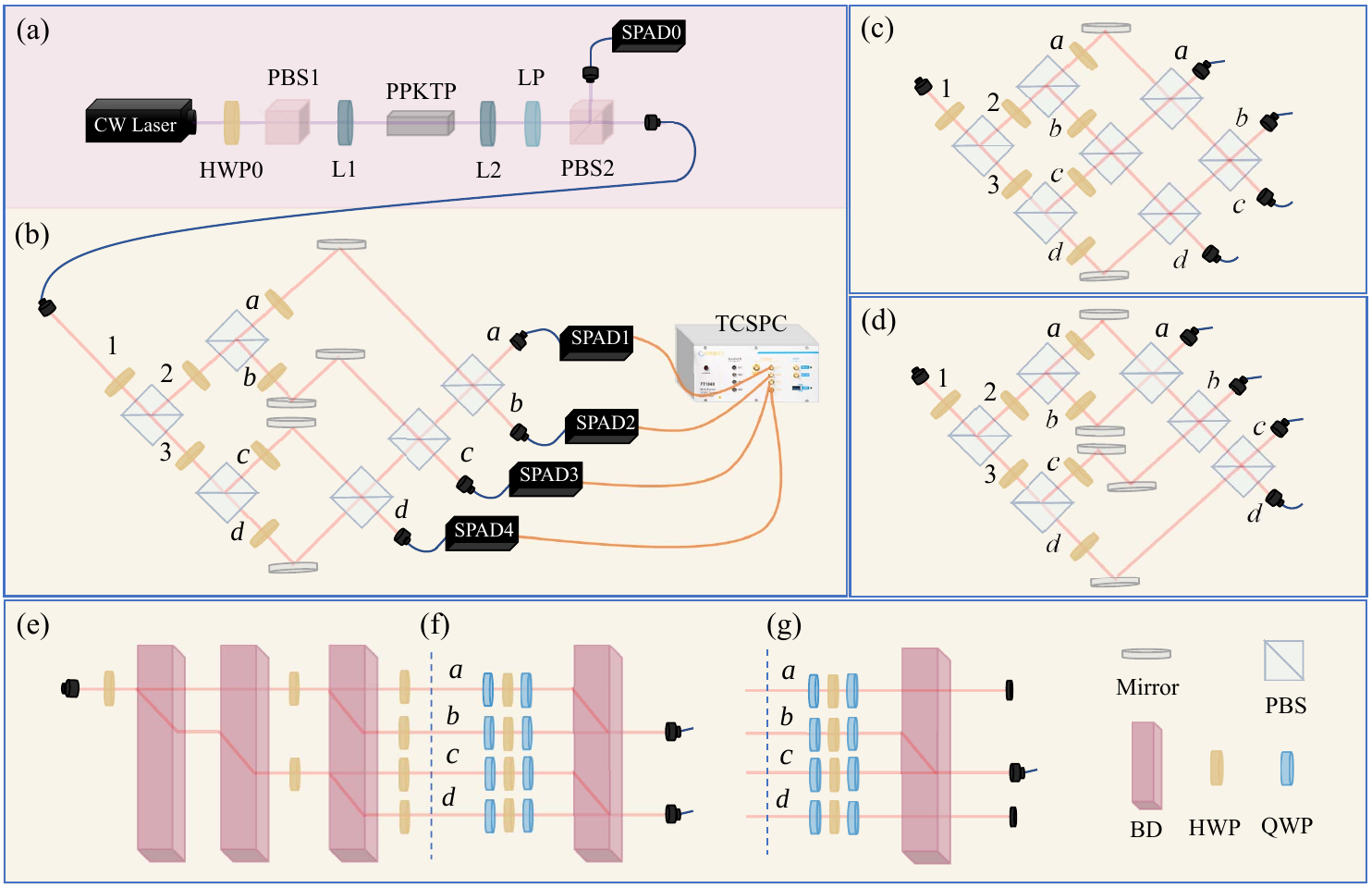}
\caption{Schematic of the experimental setup for realization of the generalized four-dimensional Pauli $X_4$ gate, Pauli $Z_4$ gate and all of their integer powers. (a) The preparation of a heralded single-photon source.  The experimental setups for the realization of (b) $X_4$ gate, (c) $X_4^2$ gate, and (d) $X_4^\dagger$ gate, where the optical elements before the mirrors present the initial state preparation and the optical elements after the mirrors describe the gate operations. (e)-(g) The experimental setups for the realization of Pauli $Z_4$, $Z_4^2$, $Z_4^\dagger$ gates.  (e) The initial state preparation of the $Z_4$, $Z_4^2$, $Z_4^\dagger$ gates. (f)-(g) The implementation of the Pauli $Z_4$, $Z_4^2$, $Z_4^\dagger$ gates and the measurement of the phase differences in spatial modes $a$ and $b$, $c$ and $d$, $b$ and $c$. The effective coincidence window (including the jitter of the detector) is about 2 ns.} \label{setup1}
\end{center}
\end{figure*}

In this Letter, we experimentally demonstrate a universal four-dimensional quantum gate set on the single photons carrying both the polarization and the spatial mode DOFs. The gate set includes the four-dimensional Pauli $X_4$ gate, Pauli $Z_4$ gate, and all of their integer powers, which can efficiently construct any four-dimensional quantum operations. The Pauli $X_4$ gate, Pauli $Z_4$ gate, and all of their integer powers are realized with the polarization-spatial DOF of the single photons. We also realize the controlled-$X_4$ gate and all of its integer powers. Our experiments overcome the obstacles of phase instability and greatly simplify the previous works based on many cascaded interferometers \cite{babazadeh2017high,gao2019arbitrary,gao2020computer,wang2021experimental,chi2022programmable}. The experimental results show both ultra-high ($ \sim 99.5\%$) gate efficiency and fidelity, {which are above error threshold for fault-tolerant quantum computation \cite{li2015resource,tzitrin2021fault}. 

%In this Letter, we experimentally demonstrate a universal four-dimensional quantum gate set on the single photons carrying both the polarization and the spatial mode DOFs. The gate set includes the four-dimensional Pauli $X_4$ gate, Pauli $Z_4$ gate, and all of their integer powers, which can efficiently construct any four-dimensional quantum operations. The Pauli $X_4$ gate, Pauli $Z_4$ gate, and all of their integer powers are realized with the polarization-spatial DOF of the single photons. We also realize the controlled-$X_4$ gate and all of its integer powers. Our experiments overcome the obstacles of phase instability and greatly simplify the previous works based on many cascaded interferometers \cite{babazadeh2017high,gao2019arbitrary,gao2020computer,wang2021experimental,chi2022programmable}. The experimental results show both ultra-high ($ \sim 99.5\%$) gate efficiency and fidelity, {which are above error threshold for fault-tolerant quantum computation \cite{li2015resource,tzitrin2021fault}. 

\emph{High-dimensional quantum gates.}---A qudit quantum gate is described in a $d$-dimensional Hilbert space $\mathscr{H}_d$ that is spanned by a set of orthogonal bases $\{|0\rangle, |1\rangle, \ldots, |d-1\rangle\}$.  The most important $d$-dimensional quantum gates are the generalized single-qudit Pauli $X_d$  gate, Pauli $Z_d$ gate, two-qudit controlled-$X_d$ gate, and all their integer powers. These gates are universal and they can construct arbitrary high-dimensional unitary transformations \cite{wang2020qudits,babazadeh2017high}. The transformations of the $d$-dimensional single-qudit $n$ (an integer number) powers of Pauli $X_d$ ($X^n_d$) gate and $n$  powers of $Z_d$ ($Z^n_d$) gate on the $d$-dimensional quantum state are expressed by \cite{bartlett2002quantum}
%
%\begin{align}
%\begin{split}         \label{eq1}
%X^n_d=\sum_{l=0}^{d-1}|l\oplus n\rangle_{\text{mod}d}\langle l|,
%\end{split}
%\end{align}
%
%
\begin{align}
\begin{split}         \label{eq1}
X^n_d|l\rangle=|l\oplus n\rangle_{\text{mod}d}, \qquad Z^n_d|l\rangle=\omega^{n\cdot l}|l\rangle.
\end{split}
\end{align}
Here $l \in \{0, 1, \dots, d-1\}$, $|l\oplus n\rangle_{\text{mod}d}$=$(l+n)$ modulo $d$, and $\omega$=exp($2\pi i/d$). The $X^n_d$ gate is a cyclic operation in which each quantum state is transformed to its $n$-th nearest state in a clockwise direction. The $Z^n_d$ gate is a phase operation in which each quantum state is introduced a state-dependent phase. The generalized $n$ powers of Pauli $Y_d$ ($Y^n_d$) gate can be given by $Y^n_d=X^n_dZ^n_d$. When $n=1$ and $d=2$, they would simplify to qubit Pauli $X$ gate, Pauli $Y$ gate, and Pauli $Z$ gate. Besides, the important two-qudit gate is the controlled-$X^n_d$ ($CX^n_d$) gate, which is formulated as \cite{wang2020qudits}
\begin{align}
\begin{split}         \label{eq2}
CX^n_d(|k\rangle|l\rangle)=|k\rangle|k\oplus l\oplus (n-1)\rangle_{\text{mod}d}.
\end{split}
\end{align}
The $CX^n_d$ gate realizes a cyclic operation on the target qudit $|l\rangle$ by manipulating the value of the controlled qudit $|k\rangle$ and leaves it unchanged on the controlled qudit. In qubit version, the $CX^n_d$ gate becomes the well-known controlled-NOT (CNOT) gate.

\begin{figure*}   [tpb] %[!h]
\begin{center}
\includegraphics[width=17cm,angle=0]{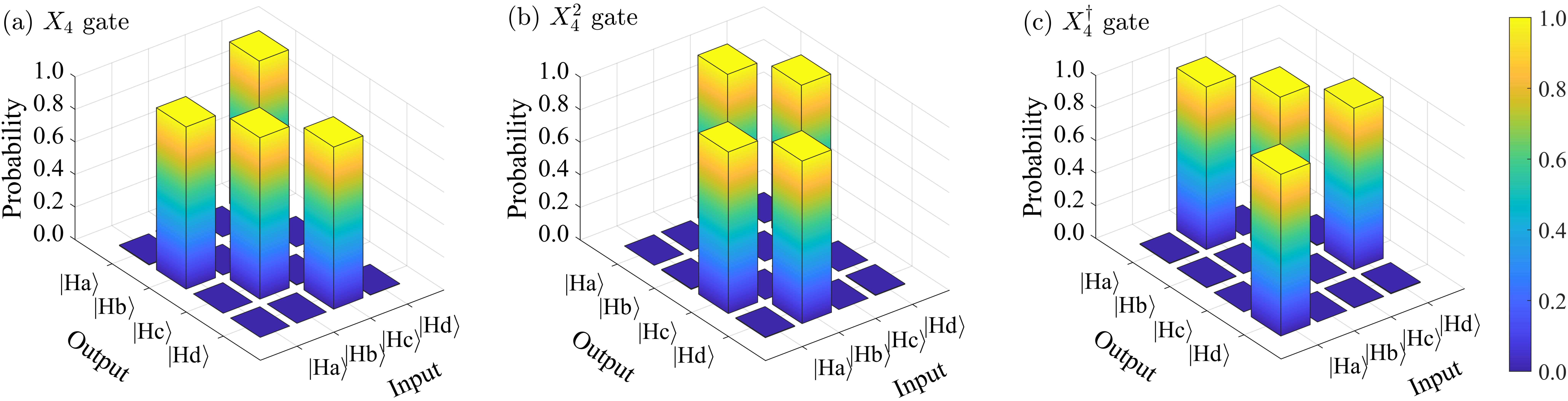}
    \caption{Truth tables for (a) $X_4$ gate, (b) $X^2_4$ gate, and (c) $X^{\dagger}_4$ gate. After preparing a qudit in one of the four input computational basis from $|Ha\rangle$ to $|Hd\rangle$, the probabilities of all output basis states are measured in 10s. The average conversion efficiencies of truth tables for $X_4$ gate, $X^2_4$ gate and $X^{\dagger}_4$ gate are 0.9941, 0.9961, and 0.9950, respectively.} \label{Truth1}
\end{center}
\end{figure*}

\emph{Experimental setup.}---Quantum gates $X_4$, $X^2_4$, $X^\dagger_4$ ($X^\dagger_4$=$X^3_4$) and $Z_4$, $Z^2_4$, $Z^\dagger_4$ ($Z^\dagger_4$=$Z^3_4$) are sufficient to construct arbitrary quantum operations in the four-dimensional space. The experimental setup for the realization of these elementary gates is shown in Fig. \ref{setup1}, where the gate qudit is encoded on the polarization-spatial DOF of the single photons, i.e., $|0\rangle\leftrightarrow|Ha\rangle$, $|1\rangle\leftrightarrow|Hb\rangle$, $|2\rangle\leftrightarrow|Hc\rangle$, and $|3\rangle\leftrightarrow|Hd\rangle$. The horizontally $H$-polarized DOF of photons controls the spatial mode DOF ($a$, $b$, $c$, $d$) of photons to route the photons into the corresponding logic gates by using some linear optical elements.

% We encoded qudits in the spatial degree of freedom of the photons and utilize the polarization degree of freedom of the photons to exert control whether the photon passes through I gate or X gate.

Figure \ref{setup1}(a) illustrates a heralded single-photon source, in which a continuous-wave diode laser (CW Laser) is employed to generate a pump laser beam with a central wavelength of 405 nm and an output power of 20 mW. This pump laser is utilized for the production of photon pairs at a wavelength of 810 nm through type-II spontaneous parametric down-conversion in a periodically poled potassium titanyl phosphate (PPKTP) crystal. A half-wave plate
(HWP0) and a polarization beam splitter (PBS1) are used to regulate optical power, and two lenses (L1 and L2) placed before and after the PPKTP crystal are used to focus and collimate beams. Then, the photon pairs are filtered at a long pass filter (LP) to eliminate any residual pumped laser light and they finally are split at PBS2. During this process, one photon from each pairs is detected at single-photon avalanche photodiode (SPAD0) to serve as a herald idler photon, and the other photon from pairs as signal photon $H$ is injected into a four-dimensional Pauli $X_4$ gate in Fig. \ref{setup1}(b).

As shown in Fig. \ref{setup1}(b), the $|H\rangle$ photon firstly goes through a HWP in spatial mode 1 to evolve the $H$-polarized photon into the superposition of the horizontally $H$-polarized photon and vertically $V$-polarized photon. Then a PBS reflects the $V$-polarized photon into the spatial mode $2$ and transmits the $H$-polarized photon into the spatial mode $3$. Subsequently, the photons in spatial modes $2$ and $3$ go through HWPs, PBSs, and HWPs respectively, in which the PBSs divide the photons into the spatial modes $a$, $b$, $c$, $d$. Because the parameters of the HWPs can be adjusted arbitrarily, the four-dimensional Hilbert space is spanned by the trajectories of photons in $a$, $b$, $c$, and $d$, which generates a general superposition of polarization-spatial  states (see Supplementary Materials)
\begin{align}
\begin{split}         \label{eq3}
\alpha |Ha\rangle+\beta|Hb\rangle+\gamma|Hc\rangle+\delta|Hd\rangle.
\end{split}
\end{align}
Here the complex coefficients $\alpha$, $\beta$, $\gamma$, and $\delta$ satisfy the normalization condition $|\alpha|^2$+$|\beta|^2$+$|\gamma|^2$+$|\delta|^2$= 1. After the initial state is prepared, the photon is routed to a $X_4$ gate that is composed of three PBSs, which evolves the initial state in Eq. \eqref{eq3} as
\begin{align}
\begin{split}         \label{eq4}
\alpha |Hb\rangle+\beta|Hc\rangle+\gamma|Hd\rangle+\delta|Ha\rangle.
\end{split}
\end{align}
From Eq. \eqref{eq3} to Eq. \eqref{eq4}, one can see that the $X_4$ gate is accomplished. In this way, the $X^2_4$ gate and $X^\dagger_4$ gate also can be realized by routing the photon to different spatial modes using the PBSs, and the corresponding experimental setups are presented in Fig. \ref{setup1}(c) and Fig. \ref{setup1}(d), respectively.

Our four-dimensional  universal quantum gates  exhibit advantages over quantum walk.  In our experiment, $X$ gate can transit from the spatial
mode $d$ to the spatial
cmode $a$ using merely three PBSs. The quantum walk scheme requires three steps to complete, which is an operation that requires fifteen  PBSs and thirty HWPs.   This is primarily due to the fact that the number of optical elements required for quantum walk increases polynomially with the number of steps. In the quantum walk scheme, a $d$-dimensional Pauli $X_d$ gate and $CX_d$ gate can be realized by using $(1.5d^2-2.5d+1)$ PBSs and $(3d^2-5d+2)$ HWPs \cite{Do:05,PhysRevLett.102.180501,PhysRevA.81.042330}.

We also experimentally realize the Pauli $Z_4$, $Z_4^2$, and $Z_4^\dagger$ gates shown in Figs. \ref{setup1}(e)-(g). Specifically, Fig. \ref{setup1}(e) presents the initial state preparation using the beam displacers (BDs) and HWPs, which can modeled as a general state in Eq. \eqref{eq3}. In this case, it is equivalent to the method of using PBSs and HWPs for the initial state preparation of the $X_4$ gate. Figs. \ref{setup1}(f)-(g) show the gate operations and the relative phase measurements, where a combination of two quarter-wave plates (QWPs) and one HWP is usd to introduce the state-dependent phase operations in corresponding spatial modes, and finally measuring the relative phase between two pairwise locations by the interference of the photons from the spatial modes $a$ and $b$, $c$ and $d$, $b$ and $c$, respectively.

In experiments, the output signal photons after the gate operations are detected by a measurement device consisting of four SPADs (SPAD1, SPAD2, SPAD3, SPAD4) and a time-correlated single photon counting (TCSPC). This allows the photon number statistics in each output spatial modes to count the probabilities of all four elementary output bases for the Pauli $X_4$, $X_4^2$, $X_4^\dagger$ gates, and it also ascertains the relative phases between pairwise output spatial modes for the Pauli $Z_4$, $Z_4^2$, $Z_4^\dagger$ gates. This measurement process is sustained over a duration of 10 seconds by registering the coincidence  between the SPAD1, SPAD2, SPAD3, SPAD4 and triggering SPAD0, respectively. For each measurement, we record the detection of approximately 9000 heralded single photons by registering the clicks over a duration of 1 second.

%We calculate the gate efficiency and fidelity to evaluate the performance of the experimental gates. The gate efficiency and fidelity are defined as $P(i, j)=n_{ij}/\sum_kn_{ik}$ and $F(\rho_e,\rho_t)=\text{Tr}\big(\sqrt{\sqrt{\rho_e}\rho_t\sqrt{\rho_e}}\big)$ \cite{nielsen2010quantum}, respectively. Here $n_{ij}$ denotes the output photon number in the $j$-th spatial mode when the input photon is in the $i$-th spatial mode, and $\sum_kn_{ik}$ denotes the summation over the photon number in all possible output spatial modes when the input photon is in the $i$-th spatial mode. $\rho_e$ and $\rho_t=U\rho_{i}U^\dagger$ ($\rho_i$ is input sate and $U$ is the transformation of the gates) denote the experimental output state and the theoretical output state, respectively.  We reconstruct the truth tables for the $X_4$, $X_4^2$, and $X_4^\dagger$ gates plotted in Fig. \ref{Truth1}, which describe the population of all computational basis output states to each of the computational basis input states. We obtain the efficiency of the $X_4$, $X_4^2$, and $X_4^\dagger$ gates and list it in Tab. \ref{table1}.   The average efficiency of the $X_4$, $X_4^2$, and $X_4^\dagger$ gates is $99.41\%$, $99.61\%$, $99.50\%$, respectively.  The experimental errors stem mainly from the imperfections of the single photon source and the single photon detectors.

\emph{Experimental results.}---The conversion efficiency and gates fidelity can be used to evaluate the performance of the quantum gates. The conversion efficiency of the gate is defined as $\mathcal{P}(i, j)=n_{ij}/\sum_kn_{ik}$. Here $n_{ij}$ denotes the output photon number in the $j$-th spatial mode when the input photon is in the $i$-th spatial mode, and $\sum_kn_{ik}$ denotes the summation over the photon number in all possible output spatial modes when the input photon is in the $i$-th spatial mode.  We reconstruct the truth tables for the $X_4$, $X_4^2$, and $X_4^\dagger$ gates plotted in Fig. \ref{Truth1}, which describe the population of all computational basis output states to each of the computational basis input states. We calculate the efficiencies of the $X_4$, $X_4^2$, and $X_4^\dagger$ gates and list them in Tab. \ref{table1}.   The average efficiencies of the $X_4$, $X_4^2$, and $X_4^\dagger$ gates are $99.41\%$, $99.61\%$, $99.50\%$, respectively.  The experimental errors stem mainly from the imperfections of the single photon source and the single photon detectors.

\begin{table} [tpb]
\centering\caption{The in-out efficiency $\mathcal{P}(i, j)$ for the $X_4$, $X_4^2$, and $X_4^\dagger$ gates in our experimental setups.}
\begin{tabular}{cccccccccc}

\hline  \hline
&Input mode       &\quad & $|Ha\rangle$  &\quad & $|Hb\rangle$ &\quad & $|Hc\rangle$ &\quad  &$|Hd\rangle$   \\

\hline

& $X_4$ gate           &\quad\, & $99.58\%$  &\quad\, &  $99.13\%$   &\quad\, & $99.60\%$  &\quad\,  & $99.31\%$     \\
& $X_4^2$ gate         &\quad\, & $99.19\%$  &\quad &  $99.84\%$   &\quad & $99.96\%$  &\quad  & $99.46\%$     \\
& $X_4^\dagger$ gate   &\quad\, & $99.33\%$  &\quad &  $99.81\%$   &\quad & $99.77\%$  &\quad  & $99.11\%$     \\

\hline  \hline
\end{tabular}\label{table1}
\end{table}

In order to check the transformations are quantum gates instead of classical gates, we need to input the four-dimensional state in a quantum superposition way. In experiments, we send the photons prepared in an equal superposition state into the gate operations and then measure the output state. We calculate the gate fidelity $\mathcal{F}(\rho_{e},\rho_t)=\text{Tr}\big(\sqrt{\sqrt{\rho_e}\rho_t\sqrt{\rho_e}}\big)$ between the experimental output state $\rho_e$ and the theoretical output state $\rho_t=U\rho_{i}U^\dagger$ ($\rho_i$ is input sate and $U$ is the transformation of the gates) \cite{nielsen2010quantum}. We find that the fidelities for $X_4$, $X_4^2$, and $X_4^\dagger$ gates are $\mathcal{F}_{X_4}=99.70\%$, $\mathcal{F}_{X_4^2}=99.80\%$, and $\mathcal{F}_{X_4^\dagger}=99.75\%$, which go beyond significantly the maximum fidelity for these classical gates bounded by $\mathcal{F}_\text{cl}=49.82\%$ in our experiments \cite{babazadeh2017high}. These results suggest the gates run with ultra-high quality in a coherent way and it can also obtain the similar outcome for the other possible coherent superpositions.

\begin{figure}   [tpb] %[!h]
\begin{center}
\includegraphics[width=8cm,angle=0]{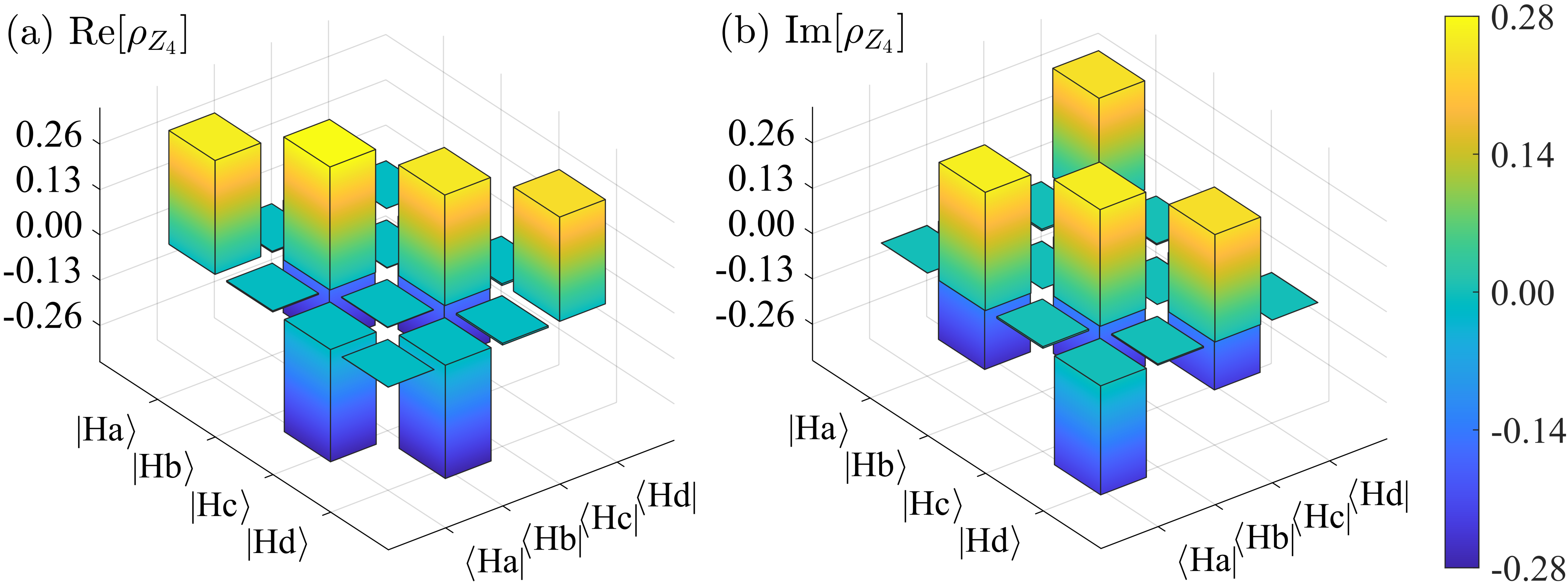}
  \caption{The reconstructed density matrix $\rho_{Z_4}$ for the $Z_4$ gate. (a) and (b) are the real part and the imaginary part of the density matrix for the $Z_4$ gate, respectively.  The fidelity of the $Z_4$ gate is 99.81\%.} \label{Z4gate}
\end{center}
\end{figure}

Figure \ref{Z4gate} shows the experimental density matrix for the $Z_4$ gate in an input equal superposition state, where the real and imaginary parts are reconstructed in Fig. \ref{Z4gate}(a) and Fig. \ref{Z4gate}(b), respectively. The density matrices for $Z_4^2$ gate and $Z_4^\dagger$ gate can be found in Supplementary Materials. We obtain the fidelity of the $Z_4$, $Z_4^2$ and $Z_4^\dagger$ gates $\mathcal{F}_{Z_4}=99.81\%$, $\mathcal{F}_{Z_4^2}=99.55\%$, and $\mathcal{F}_{Z_4^\dagger}=99.83\%$. Besides, by adjusting the angles of the HWPs for the initial state preparation in Figs. \ref{setup1}(b)-(d), we also experimentally realize the high-dimensional controlled-cyclic gates where the controlled qubits are encoded on the polarization states and the target qudits are encoded on the spatial mode states. The controlled-cyclic ($CX_4^n$) gate in an eight-dimensional hybrid Hilbert space is expressed by
\begin{align}
\begin{split}         \label{eq4}
CX^n_4=|V\rangle\langle V|\otimes I_4+|H\rangle\langle H|\otimes X_4^n.
\end{split}
\end{align}
Here $I_4$ is a four-dimensional identical operation. When the control qubit is $V$-polarized photon, the $CX_4^n$ gate preforms an $I_4$ operation and when the control qubit is $H$-polarized photon, the $CX_4^n$ gate preforms a $X^n_4$ operation. We initiate the preparation of a qudit in one of the eight elementary basis input states from $|Va\rangle$, $\ldots$, $|Vd\rangle$, $|Ha\rangle$, $\ldots$,  $|Hd\rangle$, and execute a measurement procedure to count the probabilities of all eight elementary output states. The experimental truth table of the $CX_4$ gate is presented in Fig. \ref{CX4} with an average conversion efficiency of 99.25\%. The average efficiencies of the $CX_4^2$ and $CX_4^\dagger$ gates are $99.56\%$, and $99.47\%$, and their truth tables can be found in Supplementary Materials. We also obtain the fidelities for all these gates in an equal superposition input state, which are given by $\mathcal{F}_{CX_4}=99.62\%$, $\mathcal{F}_{CX_4^2}=99.78\%$, and $\mathcal{F}_{CX_4^\dagger}=99.73\%$.

\begin{figure}   %[htpb] %[!h]
\begin{center}
\includegraphics[width=8cm,angle=0]{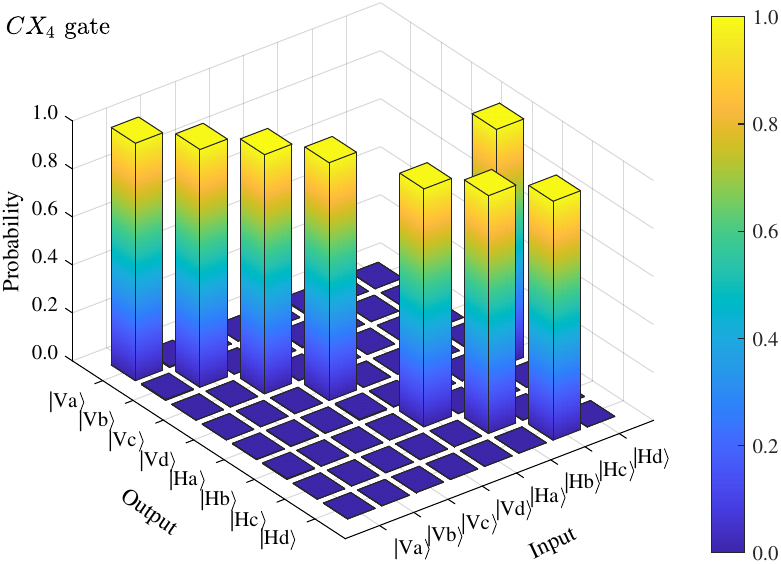}
    \caption{Truth tables for the $CX_4$ gate. The average efficiency of the $CX_4$ gate is 99.25\%.} \label{CX4}
\end{center}
\end{figure}

\emph{Conclusion.}---We have investigated the implementations of four-dimensional generalized Pauli $X_4$ gate, $Z_4$ gate, and all of their integer powers based on the polarization-spatial DOF of the single photon.  These quantum gates form a completed basis in the four-dimensional Hilbert space, which can construct arbitrary four-dimensional quantum operations. Besides, we also realized the controlled-$X_4$ gate and its all integer powers.  Our experimental setups greatly simplify the implementations of these gates and the experimental results show both the ultra-high fidelity and efficiency, which improves significantly the previous works based on OAM \cite{babazadeh2017high,gao2019arbitrary,gao2020computer,wang2021experimental}.

%\textcolor{blue}{Our four-dimensional  universal quantum gates  exhibit advantages over quantum walk.  In our experiment, $X$ gate can transit from the spatial mode $d$ to the spatial mode $a$ using merely three PBSs. The quantum walk scheme requires three steps to complete, which is an operation that requires fifteen $(4+5+6=15)$ PBSs and thirty HWPs.   This is primarily due to the fact that the number of optical elements required for quantum walk increases polynomially with the number of steps. In the quantum walk scheme, a $d$-dimensional Pauli $X_d$ gate and $CX_d$ gate can be realized by using $(1.5d^2-2.5d+1)$ PBSs and $(3d^2-5d+2)$ HWPs \cite{Do:05,PhysRevLett.102.180501,PhysRevA.81.042330}. }

These elementary high-dimensional quantum gates have important applications in many high-dimensional quantum information processing tasks, such as the preparation of high-dimensional entangled states \cite{wang2017generation}, high-dimensional quantum key distribution \cite{mafu2013higher}, high-dimensional quantum teleportation \cite{luo2019quantum}, and quantum information transfer \cite{feng2022quantum}. 
Our experiments can easily scale up more spatial modes to realize any higher-dimensional quantum operations. In this way, a $d$-dimensional Pauli $X_d$ gate and $CX_d$ gate can be realized by using $d-1$ PBSs, and a $d$-dimensional Pauli $Z_d$ gate can be realized by using $d$ HWPs and $2d$ QWPs. The work opens up a new way to construct a large-scale qudit-based photonic chip quantum processor with a polynomial resource cost \cite{chi2022programmable}. 
%Our experiments can easily scale up more spatial modes to realize any higher-dimensional quantum operations. In this way, a $d$-dimensional Pauli $X_d$ gate and $CX_d$ gate can be realized by using $2d-1$ HWPs and $2(d-1)$ PBSs, and a $d$-dimensional Pauli $Z_d$ gate can be realized by using $5d-1$ HWPs and QWPs and $d$ BDs. The work opens up a new way to construct a large-scale qudit-based photonic chip quantum processor with a polynomial resource cost \cite{chi2022programmable}. 

This work is supported by   National Natural Science Foundation of China (Grant No.92365115), Beijing Institute of Technology Research Fund Program for Young Scholars.

% our proposals only need some common linear optical elements, but do not need the multiple parity sorters or multiple cascaded interferometers, that allows the experimental implementation with the high phase stability during operation and robust fidelity.

%\bibliography{mybibliography}
%merlin.mbs apsrev4-1.bst 2010-07-25 4.21a (PWD, AO, DPC) hacked
%Control: key (0)
%Control: author (8) initials jnrlst
%Control: editor formatted (1) identically to author
%Control: production of article title (-1) disabled
%Control: page (0) single
%Control: year (1) truncated
%Control: production of eprint (0) enabled
%

\end{document}

\title{Supplementary Materials for ``Experimental realization of universal high-dimensional quantum gates with ultra-high fidelity and efficiency"}

\author{Zhe Meng}\thanks{These two authors contributed equally to this work.}
\affiliation{Center for Quantum Technology Research and Key Laboratory of Advanced Optoelectronic Quantum Architecture and Measurements (MOE), School of Physics, Beijing Institute of Technology, Beijing 100081, China}	
\author{Wen-Qiang Liu}\thanks{These two authors contributed equally to this work.}
\affiliation{Center for Quantum Technology Research and Key Laboratory of Advanced Optoelectronic Quantum Architecture and Measurements (MOE), School of Physics, Beijing Institute of Technology, Beijing 100081, China}	

\author{Bo-Wen Song}
\affiliation{Center for Quantum Technology Research and Key Laboratory of Advanced Optoelectronic Quantum Architecture and Measurements (MOE), School of Physics, Beijing Institute of Technology, Beijing 100081, China}	

\author{Xiao-Yun Wang}
\affiliation{Center for Quantum Technology Research and Key Laboratory of Advanced Optoelectronic Quantum Architecture and Measurements (MOE), School of Physics, Beijing Institute of Technology, Beijing 100081, China}

\author{An-Ning Zhang}
\email{Corresponding author. Anningzhang@bit.edu.cn}
\affiliation{Center for Quantum Technology Research and Key Laboratory of Advanced Optoelectronic Quantum Architecture and Measurements (MOE), School of Physics, Beijing Institute of Technology, Beijing 100081, China}

\author{Zhang-Qi Yin}
\email{Corresponding author. zqyin@bit.edu.cn}
\affiliation{Center for Quantum Technology Research and Key Laboratory of Advanced Optoelectronic Quantum Architecture and Measurements (MOE), School of Physics, Beijing Institute of Technology, Beijing 100081, China}

\date{\today }

\renewcommand{\thefigure}{S\arabic{figure}}
\renewcommand{\theequation}{S\arabic{equation}}

%\pacs{03.67.Lx, 42.50.Ex, 05.70.Ln}

\maketitle

%\section{Introduction}

%\newpage

\section{The proposals for the realizing the four dimensional $X_4$ gate, $Z_4$ gate, and all of their integer powers.}

\begin{figure}   [htpb] %[!h]
\begin{center}
\includegraphics[width=14cm,angle=0]{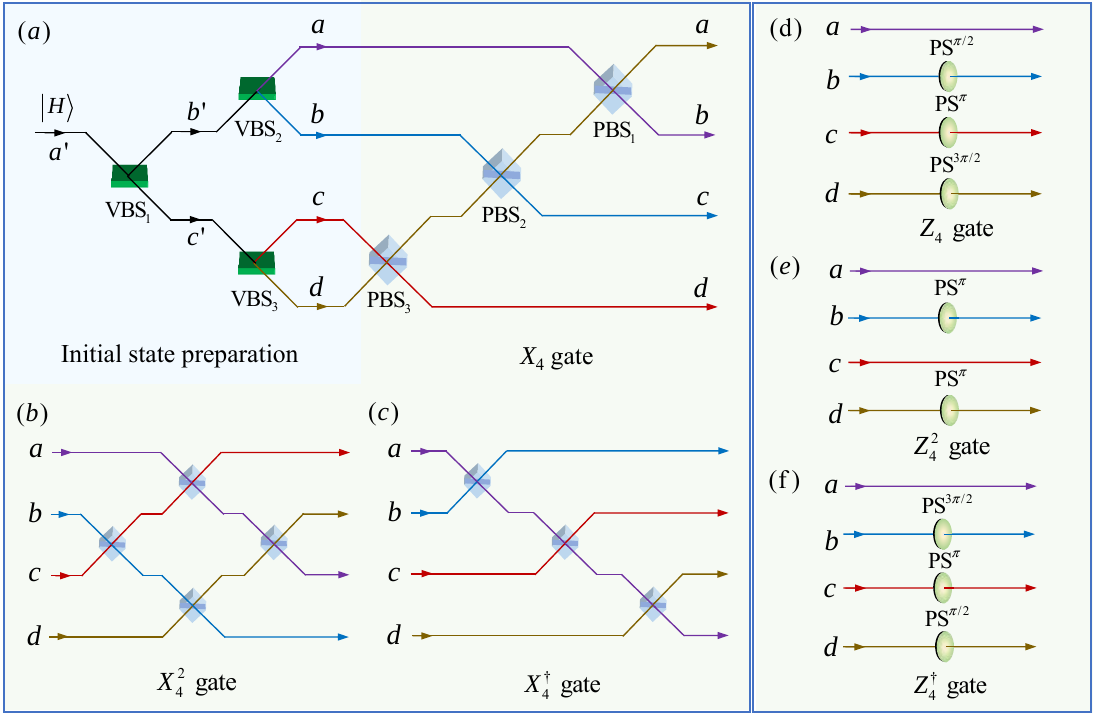}
\caption{Schematic proposals for realizing the four dimensional Pauli $X_4$ gate, $Z_4$ gate, and all of their integer powers. (a) The initial state preparation and the realization of the $X_4$ gate. A horizontally polarized photon $|H\rangle$ is injected in the spatial mode $a'$ to prepare a general initial state by using three spatial variable beam splitters (VBSs). Then the photon is routed to a $X_4$ gate composed of three polarized beam splitters (PBSs). The PBSs change the spatial mode of the incident photon from ($|Ha\rangle$, $|Hb\rangle$, $|Hc\rangle$, $|Hd\rangle$) to ($|Hb\rangle$, $|Hc\rangle$, $|Hd\rangle$, $|Ha\rangle$), because each PBS transmits the $H$-polarized photon and reflects the vertically $V$-polarized photon. (b) The realization of the $X^2_4$ gate where the spatial mode of the photon is changed from ($|Ha\rangle$, $|Hb\rangle$, $|Hc\rangle$, $|Hd\rangle$) to ($|Hc\rangle$, $|Hd\rangle$, $|Ha\rangle$, $|Hb\rangle$). (c) The realization of the $X^\dagger_4$ gate where the spatial mode of the photon is changed from ($|Ha\rangle$, $|Hb\rangle$, $|Hc\rangle$, $|Hd\rangle$) to ($|Hd\rangle$, $|Ha\rangle$, $|Hb\rangle$, $|Hc\rangle$). (d) The realization of the $Z_4$ gate. (e) The realization of the $Z^2_4$ gate. (f) The realization of the $Z^\dagger_4$ gate. PS$^\theta$ denotes a phase shifter with the angle $\theta$, which introduces a mode-dependent phase $e^{i\theta}$ in the corresponding spatial modes.} \label{Fig1}
\end{center}
\end{figure}

Figure. \ref{Fig1} shows the theoretical proposals for realizing four-dimensional Pauli $X_4$ gate, $Z_4$ gate, and all of their integer powers by encoding the gate qudit on the polarization-spatial degree of freedom of the single photons, i.e., $|0\rangle\leftrightarrow|Ha\rangle$, $|1\rangle\leftrightarrow|Hb\rangle$, $|2\rangle\leftrightarrow|Hc\rangle$, and $|3\rangle\leftrightarrow|Hd\rangle$. Here $H$ is horizontally polarized state of photons and $a$, $b$, $c$, $d$ are the spatial modes of the photon.  As shown in Fig. \ref{Fig1}(a), a horizontally polarized photon $|H\rangle$ is injected into the spatial mode $a'$ to prepare the initial state. The $|Ha'\rangle$ photon firstly goes through a variable beam splitter (VBS$_1$) to be divided into two arms $b'$ and $c'$, which evolves the photon state as
\begin{align}
\begin{split}         \label{A1}
|Ha'\rangle\rightarrow r_1 |Hb'\rangle+t_1|Hc'\rangle.
\end{split}
\end{align}
Here the coefficients $r_1$ and $t_1$ are determined by the reflectivity and the transmittance of the VBS$_1$, respectively, and they satisfy $|r_1|^2+|t_1|^2=1$. The photons in spatial modes $b'$ and modes $c'$ are transmitted and reflected at VBS$_2$ and VBS$_3$ to span a four-dimensional Hilbert space in spatial modes $a$, $b$, $c$, and $d$. The VBS$_2$ and VBS$_3$ change the state in Eq. \eqref{A1} as
\begin{align}
\begin{split}         \label{A2}
r_1 |Hb'\rangle+t_1|Hc'\rangle \rightarrow  r_1t_2 |Ha\rangle+r_1r_2|Hb\rangle+t_1r_3 |Hc\rangle+t_1t_3|Hd\rangle.
\end{split}
\end{align}
Here the coefficients $r_2$, $r_3$,  $t_2$, and $t_3$ are determined by the reflectivity  and the transmittance of the VBS$_2$ and VBS$_3$, respectively, and they satisfy $|r_2|^2+|t_2|^2=|r_3|^2+|t_3|^2=1$. Because the reflectivity and the transmittance of the VBS$_1$, VBS$_2$, and VBS$_3$ can be adjusted arbitrarily \cite{reck1994experimental}, we denote $\alpha=r_1t_2$, $\beta=r_1r_2$, $\gamma=t_1r_3$, and $\delta=t_1t_3$, then Eq. \eqref{A2} is rewritten as %
\begin{align}
\begin{split}         \label{eq3}
|\varphi_1\rangle=\alpha|Ha\rangle+\beta|Hb\rangle+\gamma|Hc\rangle+\delta|Hd\rangle.
\end{split}
\end{align}
Therefore, the initial state preparation is completed. In the Figs. 1(b)-(d) of the experimental setups in the main text, we realize the same functionality as VBS by using a combination of HWP, PBS, and HWP. The HWPs positioned before the PBS serve to modulate the polarization according to coefficients $r_1$ and $t_1$. This modulation is mathematically represented as $|H\rangle\rightarrow\cos{2\theta} |H\rangle+\sin{2\theta}|V\rangle$, $|V\rangle\rightarrow \sin{2\theta}|H\rangle-\cos{2\theta}|V\rangle$, where $\theta$  signifies the angle of the optical axis of the HWP. Given that the PBS reflects the $V$-polarized photon and transmits the $H$-polarized photon, it can differentiate between the spatial modes of photons based on the ratio of $H$ and $V$ components. The HWP positioned after the PBS then transforms the polarization states of different spatial modes into the same $H$ polarization, thereby fulfilling the role of a VBS.

The complex coefficients $\alpha$, $\beta$, $\gamma$, and $\delta$ are determined by the  reflectivity and the transmittance of the VBSs and satisfy the normalization condition $|\alpha|^2$+$|\beta|^2$+$|\gamma|^2$+$|\delta|^2$= 1. After the initial state is prepared, the photon is routed to a $X_4$ gate that is composed of three polarized beam splitters (PBS$_1$, PBS$_2$, and PBS$_3$). Because the PBS transmits $H$-polarized photon and reflects vertically $V$-polarized photon, the PBS$_1$, PBS$_2$, and PBS$_3$ evolve the initial state as
\begin{align}
\begin{split}         \label{eq4}
|\varphi_1\rangle\rightarrow \alpha |Hb\rangle+\beta|Hc\rangle+\gamma|Hd\rangle+\delta|Ha\rangle=|\varphi_{X_4}\rangle.
\end{split}
\end{align}
From Eq. \eqref{eq3} to Eq. \eqref{eq4}, one can see that Fig. \ref{Fig1}(a) realizes a  $X_4$ gate. In this way, the $X^2_4$ gate and $X^\dagger_4$ gate also can be realized by routing the photon to corresponding spatial modes using the PBSs, and the corresponding proposals are presented in Fig. \ref{Fig1}(b) and Fig. \ref{Fig1}(c), respectively.

The proposals for realizing the $Z_4$, $Z^2_4$, $Z^\dagger_4$ gates are shown in Figs. \ref{Fig1}(d)-(f), respectively. In order to realize these mode-dependent phase gates, we set some phase shifters (PS$^\theta$) rotated to an angle $\theta$ in the spatial modes to introduce a relative phase $e^{i\theta}$. As shown in Fig. \ref{Fig1}(d), three phase shifters with the angles $\frac{\pi}{2}$, $\pi$, and $\frac{3\pi}{2}$ are placed respectively in the modes $b$, $c$, and $d$ to realize the $Z_4$ gate. After the photon goes through three PSs, the initial state in Eq. \eqref{eq3} is changed as
\begin{align}
\begin{split}         \label{eq5}
|\varphi_1\rangle\rightarrow \alpha |Hb\rangle+ i\beta|Hc\rangle-\gamma|Hd\rangle-i \delta |Ha\rangle=|\varphi_{Z_4}\rangle.
\end{split}
\end{align}
In this way, the $X^2_4$ gate and $X^\dagger_4$ gate also can be realized  and they are presented in Fig. \ref{Fig1}(e) and Fig. \ref{Fig1}(f), respectively. In the Figs. 1(f)-(g) of the experimental setups in the main text, we use the sandwich structure of the wave plates group QWP1-HWP1-QWP2 to realize the PSs. The angles of the wave plates group required to achieve PS are presented in Tab. \ref{table:1}.

\begin{table}[h]
\centering\caption{The angles of the wave plates group to achieve the relative phase in our experimental setups.}
\begin{tabular}{cccc}
\hline  \hline

\textbf{phase $e^{i\theta}$} & \textbf{QWP 1} & \textbf{HWP1} & \textbf{QWP2} \\
\hline
$\theta=0$ & $0^{\circ}$ & $0^{\circ}$ & $0^{\circ}$ \\

$\theta=\pi/2$ & $90^{\circ}$ & $0^{\circ}$ & $0^{\circ}$ \\

$\theta=\pi$ & $0^{\circ}$ & $90^{\circ}$ & $0^{\circ}$ \\

$\theta=3\pi/2$ & $90^{\circ}$ & $90^{\circ}$ & $0^{\circ}$ \\
\hline \hline
\end{tabular}
\label{table:1}
\end{table}

\section{The reconstructed density matrices for the $Z_4^2$ gate and $Z_4^\dagger$ gate.} \label{AppendixA}

\begin{figure*}   [tpb] %[!h]
\begin{center}
\includegraphics[width=13cm,angle=0]{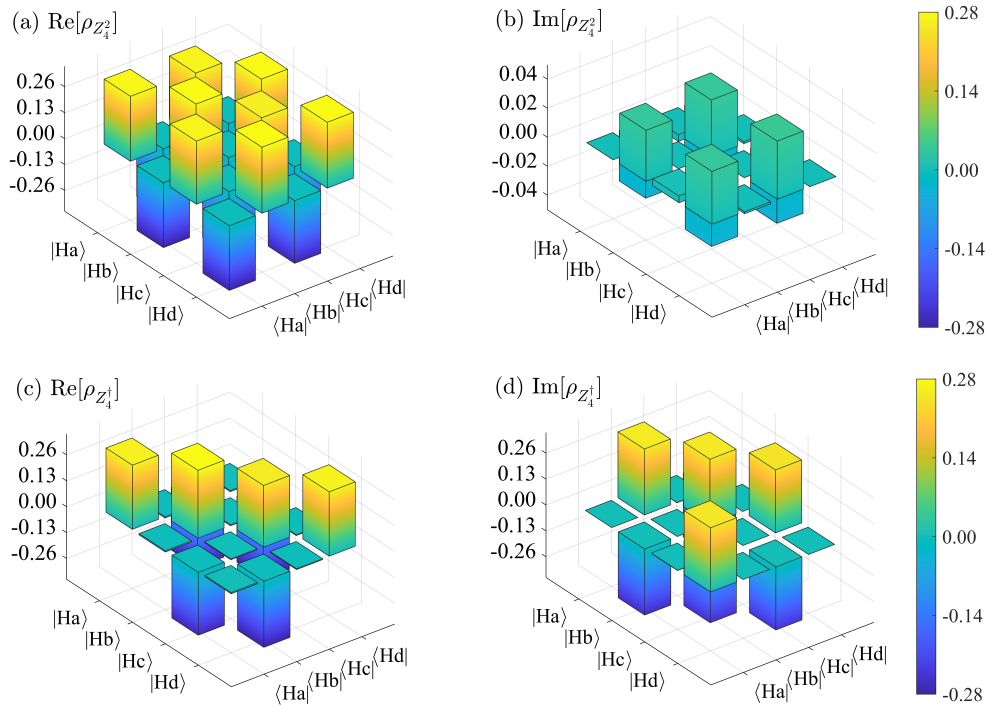}
  \caption{The reconstructed density matrices $\rho_{Z_4^2}$ and $\rho_{Z_4^\dagger}$ for the $Z_4^2$ gate and $Z_4^\dagger$ gate, respectively. (a) and (b) are the real part and the imaginary part of the density matrix for the $Z_4^2$ gate, respectively. (c) and (d) are the real part and the imaginary part of the density matrix for the $Z_4^\dagger$ gate, respectively.  The fidelities of the $Z_4^2$ gate and $Z_4^\dagger$ gate are 99.55\% and 99.83\%, respectively.} \label{Z2gate}
\end{center}
\end{figure*}

In order to assess the relative phase imparted by the $Z_4^2$ gate and $Z_4^\dagger$ gate, we introduce an initial state $|\varphi\rangle=1/2(|Ha\rangle+|Hb\rangle+|Hc\rangle+|Hd\rangle)$. We subsequently perform interferometry measurements on the spatial modes in pairs, such as $a$ and $b$, $c$ and $d$, $b$ and $c$. The relative phase bewteen two spatial modes is measured by the probability $P=1/2(1+\cos\theta)$ obtained from these measurements. In this way we measure the experimental coefficients $\alpha$, $\beta$, $\gamma$, $\delta$ by the interference of photons in spatial modes $a$ and $b$, $c$ and $d$, $b$ and $c$ for a input equal superposition state and plot the reconstructed density matrices for the $Z_4^2$ gate and $Z_4^\dagger$ gate in Fig. \ref{Z2gate}. The fidelities of the $Z_4^2$ gate and $Z_4^\dagger$ gate are 99.55\% and 99.83\%, respectively.

\section{The performance of the $CX_4$ gate, $CX_4^2$ gate, and $CX_4^\dagger$ gate.} \label{AppendixB}

\begin{figure}   %[htpb] %[!h]
\begin{center}
\includegraphics[width=15cm,angle=0]{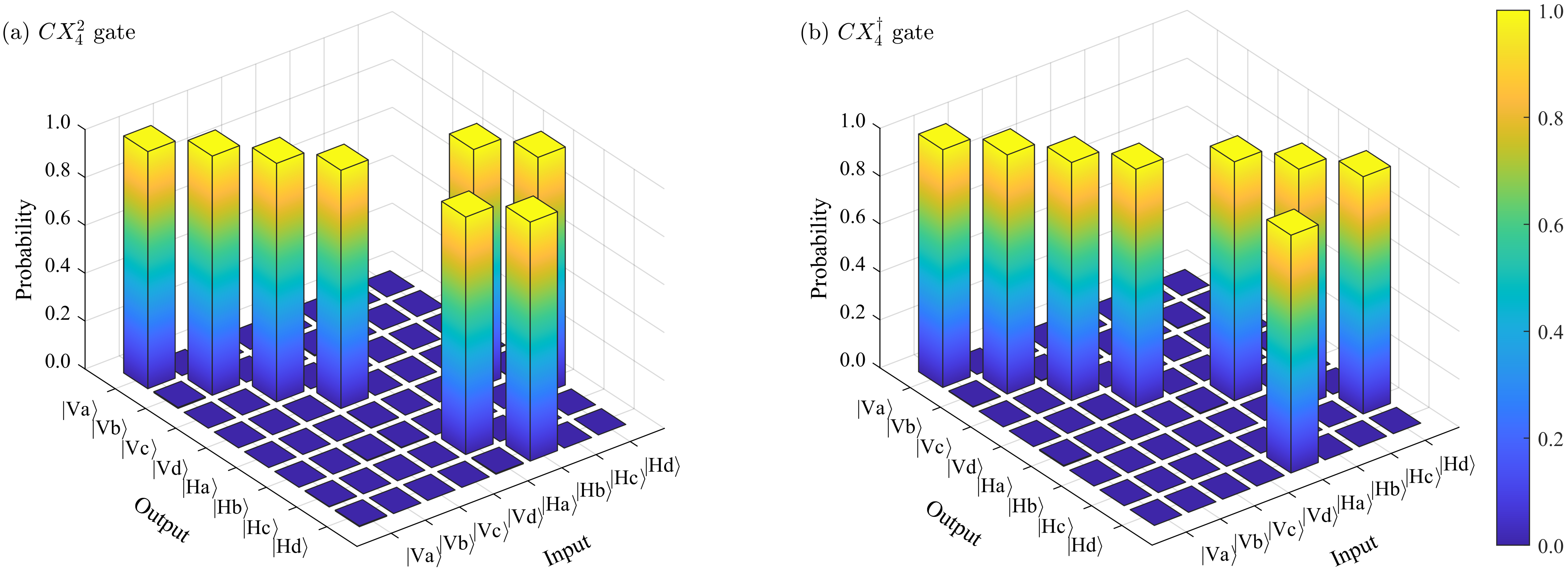}
    \caption{Truth tables for the (a) $CX_4^2$ gate and (b) $CX_4^\dagger$ gate. The average efficiencies of the $CX_4^2$ gate and $CX_4^\dagger$ gate are 99.78\% and 99.73\%, respectively.} \label{CX2}
\end{center}
\end{figure}

We check the conversion efficiencies of  the $CX_4$ gate, $CX_4^2$ gate, and $CX_4^\dagger$ gate by preparing a qudit in one of the eight input computational basis $|Va\rangle$, $\ldots$, $|Vd\rangle$, $|Ha\rangle$, $\ldots$,  $|Hd\rangle$. The probabilities of all output basis states are measured in 10s. The results of the conversion efficiencies are presented in Tab. \ref{table2} and the truth tables of them are plotted in Fig. \ref{CX2}. The average efficiencies of the $CX_4^2$ gate and $CX_4^\dagger$ gate are 99.78\% and 99.73\%, respectively.

\begin{table*} [htpb]
\centering\caption{The in-out efficiency $\mathcal{P}(i, j)$ for the $CX_4$, $CX_4^2$, and $CX_4^\dagger$ gates in our experimental setups.}
\begin{tabular}{ccccccccccccc}

\hline  \hline
&Input mode    &\,$|Va\rangle$   &\, $|Vb\rangle$  &\,$|Vc\rangle$ &\, $|Vd\rangle$ &\,$|Ha\rangle$ &\, $|Hb\rangle$ &\,$|Hc\rangle$  &\,$|Hd\rangle$   \\

\hline

& $CX_4$ gate           &\,$98.95\%$  &\, $99.10\%$  &\,$99.70\%$  &\,  $99.10\%$   &\,$99.19\%$ &\, $99.12\%$  &\,$99.56\%$  &\, $99.30\%$     \\
& $CX_4^2$ gate        &\,$99.03\%$   &\, $99.92\%$  &\,$99.61\%$  &\,  $99.49\%$   &\,$99.19\%$ &\, $99.84\%$  &\,$99.96\%$  &\, $99.45\%$     \\
& $CX_4^\dagger$ gate  &\,$99.30\%$   &\, $99.88\%$  &\,$99.47\%$  &\,  $99.42\%$   &\,$99.32\%$ &\, $99.79\%$  &\,$99.50\%$  &\, $99.09\%$     \\

\hline  \hline
\end{tabular}\label{table2}
\end{table*}

